\documentstyle[twoside,11pt,draft]{IEEEtran}
\newcommand{\be}{\begin{equation}}
\newcommand{\ee}{\end{equation}}
\newcommand{\ben}{\begin{eqnarray}}
\newcommand{\een}{\end{eqnarray}}

\newcommand{\til}{\tilde}
\newcommand{\ov}{\overline}
\newcommand{\ra}{ \rangle}
\newcommand{\la}{ \langle}
\newcommand{\h}{{\cal{H}}}

\newcommand{\CN}{C^N}
\newcommand{\kn}{| n \rangle}
\newcommand{\bn}{ \langle n |}
\newcommand{\kf}{|f \rangle}
\newcommand{\brf}{\langle f|}
\newcommand{\kftn}{|f_{S_N} \rangle}

\newcommand{\kftnc}{|f_{S_N^\bot} \rangle}
\newcommand{\kfbn}{|f_N \rangle}
\newcommand{\kaln}{|\alpha_n \ra}
\newcommand{\baln}{\la \alpha_n|}

\newcommand{\balkk}{\la \alpha_{k+1}|}
\newcommand{\kalnt}{|\til{\alpha}_n \ra}
\newcommand{\balnt}{\la \til{\alpha}_n|}
\newcommand{\kalnts}{|\til{\alpha}_n^N \ra}
\newcommand{\balnts}{\la \til{\alpha}_n^N|}
\newcommand{\kalkkkts}{|\til{\alpha}_{k+1}^{k+1} \ra}

\newcommand{\kt}{|t \ra}
\newcommand{\bt}{\la t |} 

\newcommand{\sumni}{\sum_{n=1}^\infty}
\newcommand{\sumn}{\sum_{n=1}^N}
\newcommand{\opq}{\hat{Q}}
\newcommand{\opqt}{\hat{Q}^{\dagger}}
\newcommand{\OP}{\hat{P}}
\newcommand{\FO}{\hat{F}}
\newcommand{\FOT}{\FO^\dagger}
\newcommand{\GO}{\hat{G}}
\newcommand{\GOI}{\hat{G}^{-1}}
\newcommand{\sep}{\;\; \;\;\;;\;\;\;\;\;}
\newcommand{\spa}{\, ; \,}
\newcommand{\lam}{\lambda}
\newcommand{\ketan}{|\eta_n\ra}

\newcommand{\kphin}{|\phi_n\ra}
\newcommand{\bphin}{\la \phi_n |}
\newcommand{\sumk}{\sum_{n=1}^k}
\newcommand{\sumkk}{\sum_{n=1}^{k+1}}
\newcommand{\kalnkts}{|\til{\alpha}_n^k \ra}
\newcommand{\balnkts}{\la \til{\alpha}_n^k|}
\newcommand{\kalnkkts}{|\til{\alpha}_n^{k+1} \ra}
\newcommand{\balnkkts}{\la \til{\alpha}_n^{k+1}|}
\newcommand{\OPK}{\hat{P}_k}

\newcommand{\OPKK}{\hat{P}_{k+1}}
\newcommand{\OPKKT}{\hat{P}_{k+1}^\dagger}
\newcommand{\OPKC}{\hat{P}_k^\bot}
\newcommand{\WK}{V_k^\bot}
\newcommand{\kpsik}{|{\psi}_{k+1} \ra}
\newcommand{\bpsik}{\la {\psi}_{k+1} |}
\newcommand{\kpsikp}{ \frac{\kpsik}{||\kpsik||^2}}

\newcommand{\ii}{\hat{I}}
\newcommand{\intt}{ \lim_{T \to \infty} \int_{-T}^{T}}

\title{Recursive bi-orthogonalisation approach  and orthogonal 
projectors}
\author{L. Rebollo-Neira \\
NCRG, Aston University\\ Birmingham B4 7ET, United Kingdom}

\date{}

\begin{document}
\maketitle
\begin{abstract}
An approach is proposed which, given a  family  of linearly independent
functions, constructs the appropriate bi-orthogonal  set 
so as to represent 
the orthogonal projector operator onto the corresponding subspace.
The procedure evolves iteratively and it is
endowed with the following properties: i) it yields the
desired bi-orthogonal functions avoiding the need of
inverse operations ii) it allows to quickly update a whole
family  of bi-orthogonal functions each time that a new member is
introduced in the given set. The approach is of particular relevance
to the approximation problem arising when a function is to be
represented as a finite linear superposition of non orthogonal waveforms.
\end{abstract}
\section{Introduction}
Mathematical modelling of physical systems 
often involves the representation of functions
as a linear superposition of waveforms. The formal theory of such  
representations is concerned with completeness and expansion properties of the
corresponding waveforms. The most general setting  for this purpose being
provided by the {\it {frame theory}}. Although introduced early in the
context of non-harmonic Fourier analysis \cite{du,yo} it is  only  recently
that the theory of frames has received great interest, as  
it was adopted to study wavelets and Gabor-like functions 
\cite{do1,do2,do3,wa}.
Frames are characterised by being, in general, redundant (overcomplete).
This implies that for a given frame the removal of some of its elements can still leave a frame.
Otherwise the frame is said to be an {\it {exact}} one. 
In Hilbert spaces, to  which
our considerations will be restricted, an exact frame is equivalent to a
Riesz basis. 
A further requirement on the nullity of the inner product
between different elements amounts to
an orthogonal basis.\\

The representation of functions within the frame structure
has been the subject of much work during the last decade 
\cite{do1,do2,do3,wa,ali1,ali2,ka1,ka2,han,refe,rewf,rs}. 
On the other hand, the problem of approximating functions through the linear 
superposition of a {\it {finite}} number of frame elements has been 
addressed less extensively (some discussion in the context of 
particular applications can be found in \cite{reco,rewb,red}). 
Here we would like to contribute to such a discussion by focusing on 
an approximation problem arising by considering a 
finite number of elements out of a Riesz basis. 
We believe it to be important to 
stress that approximations arising by truncating expansions given in terms of
orthogonal waveforms and those given in terms of 
non-orthogonal ones are of quite different nature.  
By truncating  an orthogonal expansion one 
obtains the approximation in the corresponding subspace which minimises the 
distance to the exact function under consideration. Nevertheless,  
 an incomplete non-orthogonal expansion does not guarantee 
an approximation of the same quality.  
If one wishes to obtain an optimal approximation in the minimum distance sense 
by means of a non-orthogonal expansion, the coefficients of the corresponding 
linear combination should be re-calculated. 
In other words,
while by truncating terms in the representation of an identity operator 
which is given  by  
orthogonal waveforms one obtains an orthogonal projector, 
an equivalent procedure in the non-orthogonal case 
does not leave an orthogonal projector.\\

Representations of the identity operator in term of 
non-orthogonal bases involves bi-orthogonal ones. Indeed, if the sequence 
$\alpha_n \spa n=1,\ldots, \infty$ is a 
Riesz basis for a Hilbert space, then there exists a reciprocal basis  
$\til{\alpha}_n \spa  n=1,\ldots, \infty$  which is bi-orthogonal to 
the former, i.e. $\la \til{\alpha}_m | \alpha_n\ra = \delta_{m,n}$. 
The identity operator in the corresponding Hilbert space  can,  thereby,  
 be expressed  as
\be
\hat{I}= \sumni \alpha_n \la \til{\alpha}_n| \,.\, \ra
\label{ioo}
\ee 
where $ \la \til{\alpha}_n| \,.\, \ra $ indicates that $\hat{I}$ acts by 
performing inner products, as in 
$\hat{I} f= \sumni \alpha_n \la \til{\alpha}_n|f \ra$. 
Now, if the sum in (\ref{ioo}) is truncated up to $N$ terms, then the 
approximation of a function $f$ that we obtain by computing 
$\sumn \alpha_n \la \til{\alpha}_n|f \ra$ is not the best approximation 
of $f$ that one can obtain as a linear superposition of $N$ elements 
$\alpha_n \spa n=1,\ldots,N$. 
This is tantamount to stating  that the operator that  is obtained by 
truncating the sum in (\ref{ioo}) up to $N$ terms is not the orthogonal 
projector onto the subspace spanned by the elements 
$\alpha_n \spa n=1,\ldots,N$. 
Indeed, as will be discussed in subsequent sections, 
the bi-orthogonality condition does not yield, per se,
orthogonal projections. In order to construct an orthogonal projector 
operator the appropriate bi-orthogonal functions need to be computed.\\

Given a set of $N$ linearly independent functions spanning a subspace, 
say $S_N$, our goal is to develop an 
effective procedure for constructing the corresponding 
bi-orthogonal functions giving rise to the orthogonal projector onto $S_N$.
An  advantage  of the algorithm we propose here  being 
its capability of building up the desired  bi-orthogonal functions avoiding 
the need of inverting operators. This feature allows to 
quickly adapt the whole set of bi-orthogonal functions each time that 
the  corresponding subspace is  enlarged by 
adding  a 
new element  to  the given initial set. We assume that all these 
elements are linearly independent, however, in the event that 
this hypothesis fails to be true, the proposed algorithm 
provides a criterion for disregarding linearly dependent elements.  
Thereby the algorithm itself can be used  for selecting subsets  
of linearly independent functions.\\

The paper is organised as follows: Section II introduces the 
notation and discusses the problem of constructing orthogonal 
projections by means of non-orthogonal waveforms. In section III 
an iterative procedure to compute bi-orthogonal functions 
giving rise to orthogonal projector operators is advanced.  
The algorithm is illustrated by computing the bi-orthogonal functions 
amounting to the orthogonal projector operator onto a subspace 
generated by a few Mexican hat wavelets. 

\section{Notation, background and preliminary considerations} 
\subsection{Bi-orthogonal expansions}
Adopting Dirac's vector notation \cite{di} we represent an 
element $f$ of a Hilbert space $\h$ as a vector $\kf$ and 
its dual as $\brf$. Given
a set of $\delta$-normalised continuous orthogonal vectors
$\{ \kt \,;\, -\infty < t < \infty \,;\, \bt t' \ra= \delta(t-t')\}$,
the unity operator in $\h$ is expressed
\be
{\ii}_{\h}= \intt \kt  \bt \; dt.
\label{i1}
\ee
Thus, for all $\kf$ and $|g \ra \in \h$, by inserting ${\ii}_{\h}$ in
$\brf  g \ra$, i.e,
\be
\brf  {\ii}_{\h} |g \ra= \intt \brf t \ra \bt g \ra\; dt
\ee
one is led to a representation of $\h$ in terms of
the space of square integrable functions, 
with $\bt g \ra= g(t)$ and $\la g \kt= \ov{ \bt g \ra}= \ov{g(t)}$, 
where $\ov{g(t)}$ indicates the complex conjugate of $g(t)$.\\ 
Let vectors $\kaln \in \h \; ;\;  n=1,\ldots, \infty$ be a Riesz Basis 
for $\h$. Then there exists a reciprocal basis  
$\kalnt \; ;\;  n=1,\ldots, \infty$ for $\h$ to which the former basis is 
bi-orthogonal i.e., $\balnt \alpha_m \ra= \delta_{n,m}$.  
These two bases amount to a representation of the identity 
operator as given by:
\be
\hat{I}= \sumni \kaln \balnt \equiv \sumni \kalnt \baln, 
\label{io}
\ee
so that every $\kf \in \h$ can be expanded in the form
\be
\kf =\sumni \kaln \balnt f \ra  \equiv \sumni \kalnt \baln f \ra.
\ee
The statements above concerning bi-orthogonal bases are well established 
results \cite{yo}. Nevertheless, the corresponding approximations arising by 
considering a finite number of vectors $\kaln \; ; \; n=1,\ldots,N$ 
have thusfar received much less consideration. It is important  
to stress that some properties holding on  truncation of  an 
orthogonal basis, no longer hold in the bi-orthogonal case. Let 
us  consider for instance that the sum in (\ref{io}) is truncated up to 
$N$ terms. Thus, rather than a representation of the identity operator 
in $\h$ we obtain an operator, $\opq$ say, given by
\be
\opq = \sumn  \kaln \balnt.
\ee
Unlike the case involving an orthogonal basis, $\opq$ is not 
the orthogonal projector operator onto the subspace 
spanned by $N$ vectors $\kaln \; ; \; n=1,\ldots,N$.  
The bi-orthogonality property of families $\kaln \; ; \; n=1,\ldots,N$
and $\kalnt \; ; \; n=1,\ldots,N$ implies that $\opq^2 =\opq$, hence 
$\opq$ is indeed a projector. However, $\opqt$ (the adjoint of $\opq$) 
is not equal to 
$\opq$ and therefore $\opq$ is not an orthogonal projector. 
As a consequence, the approximation $\kfbn$ of $\kf$ that 
we obtain from: 
\be
\kfbn= \opq f = \sumn  \kaln  \balnt f \ra   
\ee
is not the best approximation of $\kf$ that can be
obtained as a linear superposition of $N$  vectors $\kaln$. 
This being actually a major  difference 
between orthogonal and bi-orthogonal expansions.  
Whilst the condition of orthogonality  yields
orthogonal projections by truncating expansions,
the bi-orthogonality  condition, per se,
does not guarantee projections of such a nature.
If one wishes to obtain orthogonal
projections by means of bi-orthogonal families, then 
bi-orthogonal vectors specially  devised for
such a purpose must be constructed.  
\subsection{Building Orthogonal Projections}
Let us consider a subspace, $S_N$ of $\h$ spanned by 
$N$ linearly independent vectors $\kaln \;;\;n=1,\ldots,N$. 
Hence, every $\kf \in \h$ can be decomposed in the form 
\be
\kf = \kftn + \kftnc,
\ee
where $\kftn \in S_N$, i.e., $\kftn = \sumn c_n \kaln$ for some 
coefficients $c_n\;;\; n=1,\ldots,N$.
Now, in order for $ \kftn$ to be the 
orthogonal projection of $\kf$ onto $S_N$  we
should require that $\kftnc$ belongs to $S_N^\bot$, the orthogonal 
complement of $S_N$ in $\h$. This 
implies that $\kftnc$ should be orthogonal to every
vector $\kaln \;;\;n=1,\ldots,N$ and we are thus led to the following 
equations:
\be
\la \alpha_{k}\kf =  \sumn c_n \la \alpha_{k}  \kaln  \sep  k=1,\ldots,N. 
\label{ore}
\ee 
Let $\kn \;;\; n=1,\ldots,N$ be the standard basis for $\CN$, i.e,
$\la m \kn = \delta_{m,n}$ so that vector $|c \ra \in \CN$ can be 
expressed as  $|c \ra = \sumn \bn c \ra \kn= \sumn c_n \kn$ and 
equations (\ref{ore}) can be re-written  as
\be
\la \alpha_{k}\kf = \FO |c \ra
\label{or}
\ee
where  operator $\FO $  is given by
\be
\FO= \sumn \kaln \bn.
\label{of}
\ee 
Since the adjoint $\FOT$ of $\FO$ is 
\be
\FOT = \sumn \kn \baln,
\label{oft} 
\ee
the left-hand-side of equation
(\ref{or}) happens to give the components of vector $\FOT \kf \in \CN$, 
whereas on the right-hand-side we find the components of a vector $\FOT \FO
 | c \ra \in \CN$. Thus, these equations can be recast in the form
\be
\FOT \kf  = \FOT \FO  | c \ra.
\label{orf}
\ee
Since vectors $\kaln\;;\; n=1,\ldots,N$ are linearly independent 
operator $\FOT \FO$ has an inverse and $| c \ra $ is readily obtained as
\be
|c\ra = (\FOT \FO)^{-1} \FOT \kf. 
\label{la} 
\ee
The corresponding components $c_n = \bn c\ra$  being
\be
c_n = \bn c\ra = \bn  (\FOT \FO)^{-1} \FOT \kf = \balnts f \ra,
\ee 
with vectors $\balnts$ given by
\be
\balnts =  \bn  (\FOT \FO)^{-1} \FOT. 
\label{biv}
\ee 
Let us  consider now the  eigenvalue equation for the positive 
operator $\FOT \FO$ 
\be
\FOT \FO  |\eta_n \ra = \lam_n  |\eta_n  \ra\sep n=1,\ldots,N
\label{eig}
\ee
where $\lam_n >0$ and the corresponding eigenvectors $|\eta_n  \ra  \in \CN$ 
satisfy  the condition $\la \eta_k | \eta_n \ra= \delta_{k,n}$. It readily follows that the 
orthonormal vectors $|\phi_n\ra = \frac{\FO \ketan}{\sqrt{\lam_n}} \in \h $ 
are eigenvectors of the positive self-adjoint operator 
\be
\GO = \FO \FOT = \sumn \kaln \baln 
\label{go}
\ee
with associated eigenvalues 
$\lam_n \spa n=1,\ldots,N$. 
Hence, the spectral decomposition of $\GO$ is 
\be
\GO = \sumn \kphin {\lam_n}\bphin  
\ee
and that of its inverse, $\GO^{-1}$, is
\be
\GOI= \sumn \kphin  \frac{1}{\lam_n}\bphin. 
\label{esd}
\ee
Moreover, since Range($\GO$)$ \equiv S_N$, the orthogonal 
projector operator onto $S_N$ can be expressed as
\be 
\OP =  \GOI \GO =\GO \GOI=
\sumn \kphin \bphin 
\label{op}
\ee 
We are now in a position to provide an alternative form for vectors
$\balnts$  given in  (\ref{biv}).\\
{\bf{Proposition 1}:}Vectors $\balnts$, which  are
defined as
\be
\balnts= \bn (\FOT \FO)^{-1} \FOT
\ee 
can be expressed as: 
\be
\balnts = \baln \GOI 
\ee
{\bf{Proof:}} Replacing $(\FOT \FO)^{-1}$ in (\ref{biv}) by its spectral
decomposition 
$\sum_{k=1}^{N} | \eta_k\ra \frac{1}{\lam_k}\la \eta_k |$ 
we have: 
\be
\balnts=\bn (\FOT \FO)^{-1}\FOT= 
\sum_{k=1}^{N} \bn  \eta_k \ra \frac{1}{\lam_k}\la \eta_k | \FOT.
\label{pr2}
\ee
Since $ \la \eta_k | \FOT = \sqrt{\lam_k} \la \phi_k |$ and 
$|\eta_k \ra = \frac{\FOT}{\sqrt{\lam_k}} |\phi_k\ra$ from 
(\ref{pr2}) we  obtain: 
\be
\balnts= \sum_{k=1}^{N} \bn   \FOT |\phi_k\ra \frac{1}{\lam_k}
   \la \phi_k | = \bn   \FOT \GOI.
\ee
On the other hand, from (\ref{oft}) we gather that 
$\bn  \FOT = \baln$ and the proof is concluded $\Box$ \\
{\bf{Proposition 2:}} The reciprocal families 
$ \kalnts  \spa  n=1,\ldots,N$  and 
$\kaln \spa n=1,\ldots,N$  are bi-orthogonal families. \\
{\bf{Proof:}} It is straightforward by writing 
$\balnts =  \bn (\FOT \FO)^{-1} \FOT$  and 
$|\alpha_k\ra = \FO |k\ra$. Indeed, 
\be
\balnts \alpha_k\ra = \bn (\FOT \FO)^{-1} \FOT  \FO |k\ra = \bn k\ra=
\delta_{n,k}\Box
\ee 
The next proposition  shows that the families 
$\kalnts \spa n=1,\ldots,N$  and
$\kaln \spa n=1,\ldots,N$  provide us  with a representation of 
the orthogonal projection operator onto $S_N$.\\ 
{\bf{Proposition 3:}} Vectors  
$ \kalnts  \spa n=1,\ldots,N$  and
$\kaln \spa n=1,\ldots,N$ 
amount to a representation of the orthogonal projection 
operator onto $S_N$ as given by 
\be
\OP = \sumn \kaln  \balnts  \equiv \sumn \kaln \balnts
\label{opb}
\ee
{\bf{Proof:}} It immediately follows by using (\ref{go}) in (\ref{op}) 
\ben 
\OP&=&\GO \GOI = \sumn \kaln \baln \GOI = \sumn \kaln  \balnts \nonumber\\
   &=&\GOI \GO =  \sumn \GOI \kaln \baln =\sumn \kalnts \baln \Box
\een
It is clear at this point that if we introduce a
new vector $|\alpha_{N+1}\ra$ in the initial set, then
in order to represent an orthogonal projector onto the
enlarged subspace all the corresponding bi-orthogonal
vectors $|\til{\alpha}_n^{N+1} \ra\spa n=1,\ldots, N+1$
must  be re-calculated. 
The above discussion  implies that the calculation of these
vectors involves  the computation of  an inverse operator. Our goal is
to avoid the need for this calculation. In the next
section we propose a recursive procedure that enables us 
to compute bi-orthogonal vectors endowed with the desired
properties. 

\section{Recursive bi-orthogonal approach}
From the discussion of the previous section we are in a 
position to assume the existence of bi-orthogonal families of vectors 
in $\h$  giving rise to orthogonal  projections on  
the subspace they span.  Accordingly, 
given $k$ vectors $\kaln \;;\; n=1,\ldots,k$ spanning a subspace
$V_k$ let us 
denote $\kalnkts  \;;\; n=1,\ldots,k$ to the corresponding bi-orthogonal 
family  yielding a representation of the orthogonal projector 
operator onto $V_k$ i.e.,
\be
\OPK= \sumk \kalnkts \baln= \sumk \kaln \balnkts.
\label{opk} 
\ee
If we add now a new element $|\alpha_{k+1}\ra $ to the previous family, 
in order to represent the orthogonal projector onto the 
the enlarged subspace $V_{k+1}$,  we need to compute a  new 
bi-orthogonal family  $\kalnkkts \;;\; n=1,\ldots,k+1$.   
The operator $\OPKK$ in terms of the new family is:
\be
\OPKK= \sumkk \kalnkkts \baln= \sumkk \kaln \balnkkts.
\label{opkk}
\ee
Now, $V_{k+1}$ is the 
the direct sum $V_{k+1} =V_{k} \oplus |\alpha_{k+1}\ra $. Hence,
the orthogonal projector operator
onto $V_{k+1}$ can be decomposed in the fashion:     
\be
\OPKK=  \OPK + \OPKC
\label{des}
\ee
where $\OPK$ is given in (\ref{opk}) and $ \OPKC $ is the 
orthogonal projector onto the subspace $\WK $ denoting the orthogonal
complement of $V_{k}$ in $V_{k+1}$. 
Consequently, $ \WK $ is
spanned by the single element $\kpsik$, which is obtained  by
 removing from $|\alpha_{k+1}\ra$ its component in $V_{k}$, i.e.
\be
\kpsik= |\alpha_{k+1}\ra - \OPK |\alpha_{k+1}\ra. 
\label{gs}
\ee
Hence $\OPKC $ is obtained from $ \frac{\kpsik}{||\kpsik||}$
as
\be 
\OPKC =  \frac{\kpsik \bpsik}{|| \kpsik ||^2} 
\ee 
and we can explicitly  re-write (\ref{des}) in the form
\be
\sumkk \kalnkkts \baln = \sumk  \kalnkts \baln + 
\frac{\kpsik \bpsik}{||\kpsik ||^2}
\label{main}
\ee 
or, equivalently, 
\be
\kalkkkts \balkk + \sumk \kalnkkts \baln = 
\sumk \kalnkts  \baln  + \frac{\kpsik \bpsik}{||\kpsik ||^2}.
\label{main2}
\ee 
By taking the inner product of both sides of (\ref{main2}) with 
$\kalnkts \spa n=1,\ldots,k$  we  obtain:
\be
\kalkkkts \la \alpha_{k+1} \kalnkts + \kalnkkts =  \kalnkts
\sep n=1,\ldots,k.
\label{e1}
\ee 
On the other hand, by taking the inner product of both sides of 
(\ref{main2}) with 
$\kpsik$ we have
\be
\kalkkkts \la \alpha_{k+1} \kpsik = \kpsik.
\label{e2}
\ee
The last two equations yield the recurrent formula that  was 
our goal to find: 
\ben 
\kalnkkts &=&\kalnkts - 
{\kpsikp}\la \alpha_{k+1} \kalnkts 
\sep n=1,\ldots,k\nonumber\\
\kalkkkts &=& \frac{\kpsik}{\la \alpha_{k+1} \kpsik }= {\kpsikp}
\label{rec}
\een
We show now that these vectors  indeed satisfy the 
desired properties. The  next theorem is in order.\\
{\bf{Theorem 1:}} Let $\kaln \spa n=1,\ldots,k+1$ be a set of linearly 
independent vectors spanning a subspace $V_{k+1}$. Then 
vectors $\kalnkkts \spa n=1,\ldots,k+1$ constructed as prescribed in  
(\ref{rec}), by considering $|\psi_1\ra= |\alpha_1\ra $, 
satisfy the following properties:

a)are bi-orthogonal to vectors $\kaln \spa n=1,\ldots,k+1$ i.e.,
$$\la \alpha_m \kalnkkts = \delta_{m,n}$$ 
 
b)provide a representation of the orthogonal projection operator 
onto $V_{k+1}$, i.e., 
\be
\OPKK = \sumkk \kaln \balnkkts= \OPKKT =\sumkk \kalnkkts \baln
\label{teu}
\ee
Both the proofs of a) and b) are achieved by induction. The 
corresponding steps are to be found in Appendix A.\\
{\bf{Corollary 1:}} The  coefficients $c_n^{k+1}$ of the linear 
expansion 
\be
\sum_{n=1}^{k+1} c_n^{k+1} \kaln 
\ee
which approximates an arbitrary vector $\kf \in \h$ at best 
in a minimum distance sense, can be recursively obtained 
as:
\ben
c_n^{k+1}&=& c_n^{k} - \balnkts \alpha_{k+1}\ra
\frac{\bpsik f \ra}{||  \kpsik ||^2} \sep n=1,\ldots,k \nonumber\\
c_{k+1}^{k+1}&=&\frac{\bpsik f\ra}{||\kpsik||^2}, \label{c2}
\een
with 
$c_1^{1}= \frac{\la \alpha_{l_1}| f \ra}{|||\alpha_{l_1}\ra||^2}$.\\
{\bf{Proof:}} The proof  results   from  the fact that the unique 
vector in $V_{k+1}$ which minimises the distance to 
$\kf$ is obtained as $\OPKK \kf$. Indeed, let 
$|f'\ra$ be an arbitrary vector in $V_{k+1}$ and 
let us write it as $|f'\ra= |f'\ra + \OPKK \kf - \OPKK \kf$. 
If we calculate the squared distance $||\kf -  |f' \ra||^2$, 
since $(\kf - \OPKK \kf)$ is orthogonal to every vector in $V_{k+1}$, 
we have 
$$||\kf -  |f' \ra||^2 =  || \kf -  \OPKK \kf + |f'\ra -  \OPKK \kf= 
                         || \kf -  \OPKK \kf||^2 + || |f'\ra -  \OPKK \kf||^2$$
from where we gather that the distance is minimised 
if $|f'\ra \equiv \OPKK
 \kf$. Hence (\ref{c2}) readily follows by using 
(\ref{rec}) in (\ref{teu}) and identifying $c_n^{k+1}$ with $\balnkkts f \ra $
for $n=1,\ldots,k+1 \Box$\\
We illustrate the proposed approach by the
following example: Let us consider $N=5$ 
elements $|\alpha_n\ra \spa n=1,\ldots,5$
whose functional representations are given by the following 
shifted  Mexican hat wavelets
\be
\alpha_n(t)= \la t \kaln = \frac{2}{\sqrt{3} \pi^\frac{1}{4}} e^{-(t-n+1)^2}
(1-(t-n+1)^2)  \sep n=1,\ldots,5
\ee
Figure 1 plots the wavelet $\alpha_1$.  
Figure 2 plots the functions $\tilde{\alpha}_1^1$ (thin line), 
$\tilde{\alpha}_1^3$ (dotted line) and $\tilde{\alpha}_1^5$
(thick line)
which  are  involved in the representation of the  
orthogonal projectors onto the subspaces spanned, respectively, by the sets
$\{ \alpha_1\}$,  $\{\alpha_1,\alpha_2,\alpha_3\}$, 
$\{\alpha_1, \alpha_2,\alpha_3,\alpha_4,\alpha_5\}$
\section{Conclusions}
An iterative procedure, for constructing bi-orthogonal functions
yielding orthogonal projectors, has been proposed. The outcome being
relevant to  the approximation problem concerning the
representation  of a function as a linear supposition of waveforms.
The proposed  method  avoids the computation of inverse operations and allows us 
to update the whole set of bi-orthogonal functions when the
dimension of the subspace is increased. This feature is, thereby,
of great importance in situations in which the waveforms are
to be selected from a large number of possible ones \cite{relo}. \\
It is appropriate  to stress that if the available waveforms are
linearly dependent, the proposed algorithm allows one  to  select
subsets of linearly independent ones. This is to be achieved simply by 
disregarding those vector $ | \alpha_{k+1} \ra  $ yielding corresponding vectors 
$ \kpsik $ (given in (\ref{gs})) of zero norm.\\
From the above remarks it is expected that this technique should be
of assistance in a  broad range of problems concerning mathematical modelling 
of physical systems.

\section*{Acknowledgments}
Support from EPSRC (GR$/$R86355$/$01) is acknowledged.
I wish to thank to Dr S. Jain for corrections of the manuscript.

\newpage
\appendix
\section*{A. Proof of Theorem 1}
\renewcommand{\theequation}{A.\arabic{equation}}
\setcounter{equation}{0}
We show first that vectors $\kalnkkts$ 
arising from $|\psi_1 \ra = |\alpha_1 \ra $ by the 
recursive formula: 
\ben      
\kalnkkts &=&\kalnkts -
{\kpsikp}\la \alpha_{k+1} \kalnkts
\sep n=1,\ldots,k\nonumber\\
\kalkkkts &=& \kpsikp
\label{reca}
\een  
are bi-orthogonal to vectors $\kaln \;;\; n=1,\ldots,k+1$, i.e., 
they satisfy the relation 
\be
\la \alpha_m \kalnkkts = \delta_{m,n} \sep  n=1,\ldots,k+1 \sep 
m=1,\ldots,k+1
\label{bia}
\ee
For $k=0$ the relation holds because 
$|\til{\alpha}_1^1 \ra = \frac{|\alpha_1 \ra}
{|| |\alpha_1 \ra ||^2}$ and therefore $ \la \alpha_1 | \til{\alpha}_1^1 \ra =
\frac{\la \alpha_1 |\alpha_1  \ra }{ || |\alpha_1 \ra ||^2}=1$. 
Assuming that for $k+1=l$ it is true that 
\be
\la \alpha_m  |\til{\alpha}_n^l \ra = \delta_{m,n} \sep  n=1,\ldots,l \sep 
 m=1,\ldots,l
\label{hi}
\ee
we shall prove that 
\be 
\la \alpha_m |\til{\alpha}_n^{l+1} \ra = \delta_{m,n} \sep  n=1,\ldots, 
l+1 \sep m=1,\ldots,l+1.
\ee 
In order to prove this we need to discriminate  between four  
different situations concerning the values of indices $m$ and $n$. 

I)$m=1,\ldots,l$ and $n=1,\ldots,l$\\
For this situation $\la \alpha_m|\psi_{l+1} \ra =0$. Hence, using  
(\ref{reca}) and (\ref{hi}) we have:
\ben
\la \alpha_m |\til{\alpha}_n^{l+1} \ra &= &
\la \alpha_m |\til{\alpha}_n^{l} \ra - 
\frac{ \la \alpha_m |\psi_{l+1} \ra}{|||\psi_{l+1}\ra||^2} 
\la \alpha_{l+1}|\til{\alpha}_n^{l}\ra \nonumber \\
&=&  \delta_{m,n}
\een

II)$m=l+1$ and $n=1,\ldots,l$\\
Thus
\ben
\la \alpha_{l+1}|\til{\alpha}_n^{l+1}\ra &=&
\la \alpha_{l+1}|\til{\alpha}_n^{l}\ra - 
\frac{\la \alpha_{l+1} |\psi_{l+1} \ra}{|||\psi_{l+1}\ra||^2}
\la \alpha_{l+1}|\til{\alpha}_n^{l}\ra \nonumber \\ 
 &=& \la \alpha_{l+1}|\til{\alpha}_n^{l}\ra  \frac{ |||\psi_{l+1}\ra||^2 -
\la \alpha_{l+1}| \psi_{l+1}\ra }{|||\psi_{l+1}\ra||^2}=0,
\een 
since $|||\psi_{l+1} \ra ||^2 = \la \alpha_{l+1}| \psi_{l+1}\ra $.\\

III)$m=l+1$ and $n=l+1$\\
This implies\\
\be
\la \alpha_{l+1}|\til{\alpha}_{l+1}^{l+1}\ra = \frac{\la \alpha_{l+1}| \psi_{l+1}\ra}
{|||\psi_{l+1} \ra ||^2} = \frac{|||\psi_{l+1} \ra||^2}{|||\psi_{l+1} \ra||^2}=1
\ee

IV)$m=1,\ldots,l$ and $n=l+1$\\
In this case $\la \alpha_m|\psi_{l+1} \ra =0$. Hence
\be
\la \alpha_m |\til{\alpha}_{l+1}^{l+1} \ra = \frac{\la \alpha_m | \psi_{l+1} \ra}
{|||\psi_{l+1} \ra ||^2} =0
\ee
From I) II) III) and IV) we conclude that
$$\la \alpha_m \kalnkkts = \delta_{m,n} \sep n=1,\ldots,k+1 \sep 
m=1,\ldots,k+1$$ 
Now we prove that vectors $\kalnkkts \spa n=1,\ldots,k+1$ given 
in (\ref{reca})
provide a representation of the orthogonal projection operator
onto $V_{k+1}$ as given by:
\be
\OPKK=\sum_{n=1}^{k+1}\kalnkkts \baln = \OPKKT= 
\sum_{n=1}^{k+1}  \kaln \balnkkts
\label{opp}
\ee
We  prove  that $\sum_{n=1}^{k+1}\kaln \balnkkts$ 
is the orthogonal projector onto $V_{k+1}$ by proving

i)$\sum_{n=1}^{k+1}\kaln \balnkkts g \ra= |g \ra \spa \forall \,\, |g  \ra\in
V_{k+1}$ and

ii)$\sum_{n=1}^{k+1}\kaln \balnkkts  g^\bot \ra= 0 \spa
\forall \,\, |g^\bot \ra \in V_{k+1}^\bot.$\\

i)is actually a consequence of the bi-orthogonality condition
(\ref{bia}).  Because if $|g \ra\in V_{k+1}$ it can be expressed as
 $|g \ra= \sum_{n=1}^{k+1} c_n \kaln$ for some coefficients
 $c_n \spa n=1,\ldots,k+1$, hence
 $\sum_{n=1}^{k+1}\kaln \balnkkts g \ra =
  \sum_{n=1}^{k+1} c_n \kaln= |g\ra$. \\
We prove ii) by induction. For $k+1=1$ we have
$\OP_1 |g^\bot \ra=\frac{|\alpha_1^1 \ra  \la \alpha_1^1  |g^\bot \ra}
{ || |\alpha_1^1 \ra||^2}= 0 \spa\forall\,\, |g^\bot \ra \in V_1^\bot$,
since $\la \alpha_1  |g^\bot \ra=0$  for all $|g^\bot \ra \in V_1^\bot$.\\
Let us assume that
\be
\OPK |g^\bot \ra=\sum_{n=1}^{k}\kaln \balnkts g^\bot \ra = 0 \sep
\forall  \, \, |g^\bot \ra \in V_k^\bot
\label{hy}
\ee
and prove that it is then true that
\be  
\OPKK |g^\bot \ra=\sum_{n=1}^{k+1}\kaln \balnkkts  g^\bot \ra= 0  \sep \forall
\,|g^\bot \ra \in V_{k+1}^\bot.
\ee  
Indeed,
\be  
\sum_{n=1}^{k+1}\kaln \balnkkts  g^\bot \ra =
|\alpha_{k+1} \ra \frac{\bpsik g^\bot \ra}{||\kpsik||^2}
 + \OP_k |g^\bot \ra -
\OP_k |\alpha_{k+1} \ra  \frac{\bpsik g^\bot \ra}{||\kpsik||^2}=0.
\ee  
The zero value of the above equation holds  from the 
fact that $\la \alpha_n  |g^\bot \ra=0 \spa n=1,\ldots,k+1$
for all $|g^\bot \ra \in V_{k+1}^\bot$ and $\OP_k |g^\bot \ra=0$ by
hypothesis (\ref{hy}).\\
Since $\OPKK=\sum_{n=1}^{k+1}\kaln \balnkkts$ is an orthogonal projector
it is true that $\OPKK
= \OPKKT=\sum_{n=1}^{k+1}\kalnkkts \baln $ so that the proof is completed
$\Box$

%\end{references}

\newpage

\begin{figure}[h]
\begin{center}
\vspace{4cm}
\input{al1.tex}
\vspace{4cm}\\
Figure 1: Mexican hat wavelet $\alpha_1$
\end{center}
\end{figure}
\newpage

\begin{figure}[h]
\begin{center}
\input{altil.tex}
\vspace{4cm}\\
Figure 2: Functions $\tilde{\alpha}_1^1$ (thin line),
$\tilde{\alpha}_1^3$ (dotted line) and $\tilde{\alpha}_1^5$
(thick line).
\end{center}
\end{figure}

\newpage

{\bf Figure Captions}\\

{\bf Figure 1:} Mexican hat wavelet $\alpha_1$. \\

{\bf Figure 2:} Functions $\tilde{\alpha}_1^1$ (thin line),
$\tilde{\alpha}_1^3$ (dotted line) and $\tilde{\alpha}_1^5$
(thick line).

\end{document}